\documentclass[11pt,a4paper]{article}
\usepackage{natbib} 
\usepackage{amsmath,amssymb}
\usepackage{subcaption}
\usepackage{pgfplots,tikz}
\usepackage[colorlinks=false, pdfborder={0 0 0}]{hyperref}
\usepackage{cleveref} 
\usepackage[ruled,noend]{algorithm2e}
\usepackage{fullpage}

\usetikzlibrary{patterns,positioning,shapes.geometric}
\pgfplotsset{compat=1.5}
\crefname{equation}{\!\!}{\!\!}
\crefname{algocf}{Algorithm}{Algorithms} 
\Crefname{algocf}{Algorithm}{Algorithms} 
\crefname{figure}{Fig.}{Figures}
\Crefname{figure}{Figure}{Figures}
\crefname{section}{Section}{Sections}
\Crefname{section}{Section}{Sections}
\crefname{subsection}{Section}{Sections}
\Crefname{subsection}{Section}{Sections}
\crefname{appendix}{Appendix}{Appendices}
\Crefname{appendix}{Appendix}{Appendices}

\def\fixedaccent{^*}
\def\dummyelement{\phi}
\def\pr{\mathrm{pr}}
\def\LR{\Lambda}
\newcommand{\sample}[1]{\mathrm{Sample}\left(#1\right)}
\newcommand{\vM}[2]{\mathrm{vM}\left(#1,#2\right)}
\newcommand{\rvM}[2]{\mathrm{rvM}\left(#1,#2\right)}
\newcommand{\dd}{\mathrm{d}}
\newcommand{\conj}[1]{\overline{#1}}
\newcommand{\edge}[2]{\langle #1,#2\rangle}
\newcommand{\iid}{independent and identically distributed}
\newcommand{\wrt}{with respect to\ }
\newcommand{\MPPP}{\textsc{mppp}}
\DeclareMathOperator{\cd}{\mid}
\DeclareMathOperator*{\argmax}{argmax}
\DeclareMathOperator{\E}{\mathit{E}}

\title{Fingerprint Analysis with Marked Point Processes}
\author{Peter G. M. Forbes\\ University of Oxford\and Steffen Lauritzen\thanks{Corresponding author. Department of Statistics, University of Oxford, 1 South Parks Road, Oxford OX1 3TG, United Kingdom. email: steffen@stats.ox.ac.uk.}\\ University of Oxford \and Jesper M{\o}ller\\ Aalborg University}

\begin{document}
\maketitle

\begin{abstract}
We present  a framework for fingerprint matching based on marked point process models. An efficient Monte Carlo algorithm is developed to calculate the marginal likelihood ratio for the hypothesis that two observed prints originate from the same finger against the hypothesis that they originate from different fingers. Our model achieves good performance on an NIST-FBI fingerprint database of 258 matched fingerprint pairs.\\

\noindent \textbf{Keywords:} Bayesian alignment; complex normal distribution; forensic identification; likelihood ratio; marked point processes; von Mises distribution; weight of evidence.
\end{abstract}

\section{Introduction}

Fingerprint evidence has been used for identification purposes for over one hundred years. Despite this, there has been very little scientific research on the discriminatory power and error rate associated with fingerprint identification.  Within the last ten years there has been a push to move fingerprint evidence towards a solid probabilistic framework, culminating in the recent paper by \citet{RSS}.

We discuss a novel approach for fingerprint matching using marked Poisson point processes.  We develop an efficient Monte Carlo algorithm to calculate the likelihood ratio for the prosecution hypothesis that two observed prints originate from the same finger against the defence hypothesis that they originate from different fingers. \Citet{kendall} have also considered marked Poisson point process models for fingerprints, albeit for another purpose: namely, the reconstruction of fingerprint ridges from sweat pore point patterns.

\begin{figure}[ht]
        \begin{subfigure}[b]{0.3\textwidth}
                \centering
                \includegraphics[width=\textwidth]{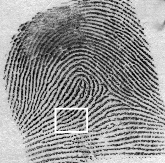}
                \caption{Exemplar fingerprint}\label{fig:FullPrint}
        \end{subfigure}
        \quad
        \begin{subfigure}[b]{0.3\textwidth}
                \centering
                \includegraphics[width=\textwidth]{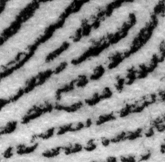} 
                \caption{Zoomed section}
        \end{subfigure}
       \quad
        \begin{subfigure}[b]{0.3\textwidth}
                \centering
                \includegraphics[width=\textwidth]{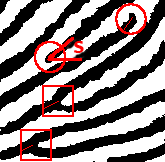} 
                \caption{Enhanced \& labelled}\label{fig:LabelledMinutiae}
        \end{subfigure} 
        \caption{A typical exemplar quality fingerprint from \citet{nistdatabase}.  The highlighted points in (c) are minutiae: circles are ridge endings and squares are bifurcations.}
        \label{fig:fingerprintstages}
\end{figure}

Fingerprint evidence is based on the similarity of two or more pictures, see \cref{fig:fingerprintstages}.  It is difficult to represent all the information from these pictures in a mathematically convenient form.  Thus most fingerprint models, including the one in \cite{RSS}, consider only a subset of the information: namely, the points on the image where a ridge either ends or bifurcates.  These points, called {minutiae},  generally contain sufficient information to uniquely identify an individual \citep{handbook,ReviewMinutiae}.  A typical full fingerprint contains 100--200 minutiae, while a low quality crime scene fingermark may contain only one dozen \citep{nistdatabase}.

\citet{RSSLauritzen} note the similarity between minutia matching and the alignment problems often studied in bioinformatics. Our model exploits ideas from the model for unlabelled point set matching in \citet{BayesianAlignmen} and applies them to the problem of fingerprint matching.  Our model could be used for an automated fingerprint identification system, or it could support a courtroom presentation of fingerprint evidence.

The paper is composed as follows. After a few preliminary specifications in \cref{sect:prelim} we develop a generic marked Poisson point process model in \cref{sect:generic} and a specific parametric version  in \cref{sect:parametric}. In \cref{sect:lrcalc} we describe our method for calculating the likelihood ratio and  in \cref{sect:datanalysis} we perform an analysis using the methodology on both simulated and real data. In the appendix we give further technical details of our computational procedures.
\section{Preliminaries and notation}
\label{sect:prelim}
\subsection{Likelihood representation of fingerprint evidence}
\label{sect:lr}

As in \cite{RSS} we discuss the situation  where we wish to compare a high-quality finger{print} $A$ taken under controlled circumstances, with a  finger{mark} $B$ found on a crime scene. We consider
two hypotheses 
\begin{align} 
	H_p&: A\mbox{ and }B\mbox{ originate from the same finger},\nonumber\\
	H_d&: A\mbox{ and }B\mbox{ originate from different fingers},
\label{eq:hypotheses} 
\end{align}
where  $H_p$ is referred to as the {prosecution hypothesis} and $H_d$ as the {defence hypothesis}. Following a tradition that goes at least back to 
\cite{lindley:77}, we follow standards in  modern evaluation of DNA   and other types of forensic evidence \citep{balding:05,aitken:taroni:04} and quantify the weight-of-evidence by calculating a {likelihood ratio} between $H_p$ and $H_d$,
\begin{equation}
\LR=\frac{\pr(A,B\cd H_p)}{\pr(A,B\cd H_d)}.
\label{eq:LR}
\end{equation}
The likelihood ratio is based on probabilistic models for the generation of the fingerprint and fingermark that shall be developed in the sequel.

\subsection{Representation of fingerprints}
\label{sect:represent}

Each minutia $m$ consists of a location, an orientation, and a type: ridge ending, bifurcation, or unobserved; see \cref{fig:LabelledMinutiae}.  We represent the location with a point in the complex plane $\mathbb C$ and the orientation  with a point on the complex unit circle $\mathbb{S}^1$.  The type is represented by a number in $\{-1,0,1\}$,  where $-1$ denotes a ridge ending, 1  a bifurcation, and 0  an unobserved type.  Thus $m$ is an element of the product space $\mathbb M = \mathbb C\times \mathbb S^1\times\{-1,0,1\}$. We let $r_m,s_m$, and $t_m$ denote the projection of $m$ onto the location space, orientation space, and type space respectively.

A fingerprint $A$ or a fingermark $B$ is represented by a finite set of elements of $\mathbb M$.  We call this representation a {minutia configuration}.  Since $A$ and $B$ are observed in arbitrary and different coordinate systems, the observed minutiae are subjected to {similarity transformations}, which consist of translations, rotations, and scalings. These can be simply represented by algebraic operations with complex numbers,
\[(r_m,s_m,t_m)\mapsto(\psi r_m + \tau,\psi s_m/|\psi|,t_m).\]

\subsection{Basic distributions}
\label{sect:basicdist}
We shall use the bivariate complex normal distribution, which describes a complex random vector whose real and imaginary parts are jointly normal with a specific covariance structure \citep{goodman:63}.  The density \wrt the Lebesgue measure is
\[\varphi_2(r;\mu,\Sigma)=\exp\{-(\conj{r-\mu})^T\Sigma^{-1}(r-\mu)\}/(\pi^2|\Sigma|),\]
where $r$ and $\mu$ are two-dimensional complex numbers, $\Sigma$ is a Hermitian positive definite $2\times 2$ complex matrix with determinant $|\Sigma|$, the overline denotes the complex conjugate, and ${}^T$ denotes the vector transpose.  The standard case of $\mu=0$ and $\Sigma$ equal to the identity matrix will be denoted $\varphi_2(r)$.  When we wish to make the two arguments explicit we will write $\varphi_2(r_1,r_2;\mu,\Sigma)$ for $r_1,r_2\in\mathbb C$. The univariate density will be denoted $\varphi(r;\mu,\sigma^2)$ where $r,\mu\in\mathbb C$ and $\sigma^2>0$, with the standard case denoted $\varphi(r)$.  

The {von Mises distribution} $\vM{\nu_0}{\kappa}$ on the complex unit circle $\mathbb S^1$ with position $\nu_0$ and precision $\kappa>0$ \citep{mardiabook} has density
\begin{equation*}
\upsilon(s; \nu_0,\kappa) = I_0(\kappa)^{-1} \exp\{\kappa \Re(s\conj{\nu_0})\}
\end{equation*}
 \wrt $\nu$, the uniform distribution on $\mathbb{S}^1$, where $\Re(z)=(z+\conj{z})/2$ is the real part of $z$. The normalization constant $I_0(\kappa)$ is the modified Bessel function of the first kind and order zero \citep[chapter 10]{specialfcnshandbook}.  The von Mises distribution can be obtained from a univariate complex Normal distribution $\varphi(s;\nu_0,2/\kappa)$ (or equivalently $\varphi(s;\kappa\nu_0/2,1)$) by conditioning on $|s|=1$.

\citet{kent:77} shows that the von Mises distribution is infinitely divisible on $\mathbb{S}^1$  and thus it makes sense to define the {root von Mises} distribution $\rvM{\nu_0}{\kappa}$ by
\[ XY \sim \vM{\nu_0}{\kappa} \mbox{ whenever } X, Y \mbox{ are independent and } X, Y \sim \rvM{\nu_0}{\kappa}.\]
 The density of the root von Mises distribution is determined by a series expansion.  We refrain from giving the details as we shall not need them.
\section{A generic marked point process model}
\label{sect:generic}

\subsection{Model specification} 

We consider the observed minutia configurations $A,B\subset \mathbb M$ as thinned and displaced copies of a latent minutia configuration.  In this paper, we use the word {latent} as a synonym for {unobservable}.  This contrasts with a common usage in fingerprint forensics where a {latent fingerprint} refers to a fingermark which is difficult to see with the naked eye, but can still be observed via specialized techniques.

Both the observed and the latent minutia configurations are modelled as marked point processes.   We assume that different fingers have independent latent minutiae configurations, whether those fingers belong to the same or different individuals.  Thus we can rephrase our two model hypotheses \cref{eq:hypotheses} as
\begin{align*}
H_p&: A\mbox{ and }B\mbox{ originate from a common latent minutia configuration }M\subset\mathbb M,\\
H_d&:  A\mbox{ and }B\mbox{ originate from independent latent minutia configurations }M,M'\subset\mathbb M.
\end{align*}

In the notation of marked point processes, each minutia $m\in\mathbb M = \mathbb C\times\mathbb S^1\times\{-1,0,1\}$ is a marked point.   The projection of $m$ onto the location space $\mathbb C$, denoted $r_m$, is called a {point} and the projection onto $\mathbb S^1\times\{-1,0,1\}$, denoted $(s_m,t_m)$, is called a {mark}.  The points form a finite Poisson point process on the complex plane with intensity function $\rho:\mathbb C\to[0,\infty)$ such that $\rho_0=\int_{\mathbb C}\rho(r)\,\dd r$ is positive and finite.  The  marks are assumed to be independently and identically distributed and independent of the points.  The marks have density $g$ \wrt the product measure $\mu=\nu\times\#$, where 
$\#$ is the counting measure on $\{-1,0,1\}$.   For the latent minutiae only the types $\{-1,1\}$ have meaning so we must insist that $g(s,0)=0$ for any $s\in\mathbb{S}^1$.  

We write the resulting {marked Poisson point process} as $M\sim\MPPP{(\rho,g)}$.   The cardinality of $M$ is Poisson distributed with mean $\rho_0$, and, conditionally on the cardinality $|M|$ of $M$, the points are \iid\, with density $\rho/\rho_0$. 

The observed fingerprint $A$ is obtained from the latent minutia configuration $M$ through three basic operations, thinning, displacement, and mapping, as follows:

\emph{A1: thinning.} Only a subset of the latent minutiae are observed, resulting in $M_{A1}=\{m\in M : I_A(m)=1\}$, where the indicators $I_A(m)$ are Bernoulli variables with success probabilities $\delta_A(r_m)$.  Here $\delta_A:\mathbb C\to[0,1]$ is a Borel function which we refer to as the selection function for $A$. We then have 
\[ M_{A1}\sim\MPPP{(\rho_{A1},g_{A1})}\mbox{  where }\rho_{A1}(r)=\rho(r)\delta_A(r),\quad g_{A1}=g.\]

\emph{A2: displacement.} The locations $r_m$ in $M_{A1}$ are subjected to additive errors $e_{m}\in\mathbb C$ with density $f_A$, the orientations $s_m$ are subjected to multiplicative errors $v_{m}\in\mathbb{S}^1$ with density $h_A$, and the types are subjected to multiplicative classification errors $c_m\in\{-1,0,1\}$ with distribution $d_A$ so that $c_m=1$ corresponds to a correct classification, $c_m=0$ to the type being unobserved, and $c_m=-1$ represents a misclassification.  This results in $M_{A2}=\{(r_m+e_{m},v_{m}s_m,c_mt_m):m\in M_{A1}\}.$ Consequently, $M_{A2}\sim\MPPP{(\rho_{A2},g_{A2})}$, where
\[\rho_{A2}(r)=f_A\ast\rho_{A1}(r)=\int_{\mathbb C} f_A(e)\rho_{A1}(r-e)\,\dd e\]
is obtained by usual convolution in $\mathbb{C}$.  The mark density is
\[g_{A2}(s,t)=\sum_{{u\in\{-1,1\}}}d_A(ut)h_A\ast g_{A1}(s,u)=\sum_{{u\in\{-1,1\}}}d_A(ut)\int_{\mathbb S^1} h_A(v)g_{A1}(s\conj{v},u)\,\dd \nu(v).\]

\emph{A3: mapping.} Finally, the marked points are  subjected to a similarity transformation to obtain 
\begin{equation}A=\{(\psi_A r_m+\tau_A,\psi_As_m/|\psi_A|,t_m) : m\in M_{A2}\},
\label{eq:mappedPointSet}
\end{equation}
with $(\tau_A,\psi_A)\in\mathbb C\times(\mathbb C\setminus\{0\})$. Thus $A\sim\MPPP{(\rho_{A3},g_{A3})}$ where $\rho_{A3}(r)=\rho_{A2}\{(r-\tau_A)/\psi_A\}/|\psi_A|^2$ and $g_{A3}(s,t)=g_{A2}(s\conj{\psi_A}/|\psi_A|,t)$.

The model for $B$ is specified analogously: $B$ is the $\MPPP$ derived from a latent minutia configuration $M'$ by three similar steps B1--B3 obtained by replacing $A$ with $B$ everywhere, i.e.\ $B\sim\MPPP{(\rho_{B3},g_{B3})}$ with intensity function and the mark density defined as above, but using a new function $\delta_B$, new indicators $I_B(m)$, new distributions $f_B,h_B,d_B$, new error terms $e'_m,v'_m,c'_m$, and new parameters $\tau_B,\psi_B$.

Finally, we make the following independence assumptions.  Under $H_d$ we have $M$ and $M'$ are \iid, while under $H_p$, $M=M'$. In both cases they have distribution $\MPPP(\rho,g)$. Conditional on $M$ and $M'$, all the variables $I_A(m),e_m,v_m,c_m$ for $m\in M$, and $I_B(m),e'_m,v'_m,c'_m$ for $m\in M'$ are mutually independent with distributions which do not depend on $M$ and $M'$. 

\subsection{Density under the defence hypothesis}
The functions $\rho,g,\delta_A,\delta_B,f_A,f_B,h_A,h_B,d_A,d_B$ depend on some set of parameters denoted $\Theta$; we describe a specific choice of these functions in \cref{sect:parametric}.  In the following we suppress the dependence on $\Theta$ for ease of presentation.

In order to obtain the densities for observed minutiae configurations we introduce  the probability distribution $\zeta=\MPPP{(\varphi,1/3)}$ as a dominating measure. Using the fact that  
\[\int_{\mathbb C} \rho_{A3}(r)\,\dd r =\int_{\mathbb C} \rho_{A2}(r)\,\dd r=\int_{\mathbb C} \rho_{A1}(r)\,\dd r =
\int_{\mathbb C} \rho(r)\delta_A(r)\,\dd r,\]
the marginal density of $A$ \wrt $\zeta$  becomes
\begin{equation}
\label{eq:PointsetsMarginalADensity}
\pr(A\cd \Theta)=c(A)\exp\left\{-\int_{\mathbb C}\rho(r)\delta_A(r)\,\dd r\right\} \prod_{a\in A}\rho_{A3}(r_a)g_{A3}(s_a,t_a),
\end{equation}
where
\[c(A)=3^{|A|}\exp(1)\prod_{a\in A}\varphi(r_a)^{-1}\]
depends only on the data,  see e.g.\ \citet[p.\ 25]{Moller}. Similarly, the density $\pr(B\cd \Theta)$ of $B$ \wrt $\zeta$ is obtained by replacing $A$ by $B$ everywhere in \cref{eq:PointsetsMarginalADensity}. 
Under $H_d$, the fingerprint $A$ and fingermark $B$ are independent and thus the density \wrt $\zeta\times \zeta$ is simply the product
\begin{multline}
\pr(A,B\cd \Theta,H_d)= c(A)c(B)\exp\left\{-\int_{\mathbb C}\rho(r)\delta_A(r)\,\dd r -\int_{\mathbb C}\rho(r)\delta_B(r)\,\dd r\right\}\\
\times \left\{\prod_{a\in A}\rho_{A3}(r_a)g_{A3}(s_a,t_a)\right\} \left\{\prod_{b\in B}\rho_{B3}(r_b)g_{B3}(s_b,t_b)\right\}.
\label{eq:likelihood-D}
\end{multline}
\subsection{Density under the prosecution hypothesis}
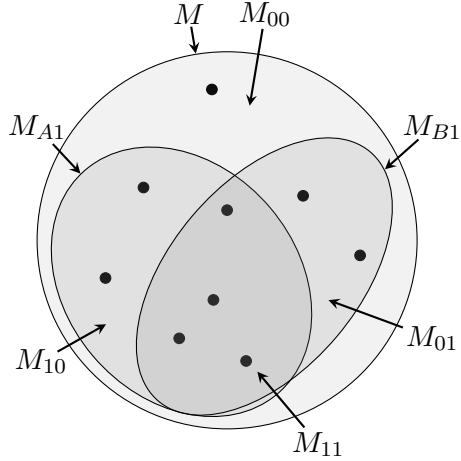
\begin{figure}[htb]
\center
        \begin{tikzpicture}[arrowlabel/.style={->,thick},>=stealth,shorten <=-1mm]
				\foreach \x/\y in {-.2/2, -1.6/-0.5,-1.1/.7, 0/.4, 1/0.59,1.75/-0.2,-.18/-.79,.25/-1.6,-.63/-1.3}
						\draw [fill=black] (\x,\y) circle (0.07cm);
			\node at (0,0) [circle,draw=black,fill=gray,fill opacity=0.1,minimum size=5cm] (M) {};				
			\node at (-.6,-.55) [shape=ellipse,draw=black,fill=gray,fill opacity=0.15,rotate=40] (A1) {\phantom{\vrule width1.9cm height2.5cm}} (A1);
			\node at (.49,-.48) [shape=ellipse,draw=black,fill=gray,fill opacity=0.15,rotate=-40] (B1) {\phantom{\vrule width1.5cm height2.8cm}};
			\node at (-.5,3) {$M$} edge[arrowlabel] (M);
			\node at (-2.5,1.5) {$M_{A1}$} edge[arrowlabel] (A1);
			\node at (2.7,1.5) {$M_{B1}$} edge[arrowlabel,shorten <= -.2cm] (B1);
			\node at (.5,3) {$M_{00}$} edge[arrowlabel,shorten >= -.7cm] (M);
			\node at (1.2,-2.7) {$M_{11}$} edge[arrowlabel,shorten >= -.4cm] (A1);
			\node at (-2.45,-1.6) {$M_{10}$} edge[arrowlabel,shorten >= -.4cm,shorten <= -.3cm] (A1);
			\node at (2.7,-1.3) {$M_{01}$} edge[arrowlabel,shorten >= -.4cm] (B1);
		\end{tikzpicture}
\caption{Partitioning the latent minutiae into those that are observed in $A$ only ($M_{10}$), $B$ only ($M_{01}$), both ($M_{11}$), and neither ($M_{00}$). The dots indicate minutiae locations.}
\label{fig:partition}
\end{figure}

The marginal densities of $A$ and $B$ are identical under both $H_d$ and $H_p$, but to obtain the joint density of $(A,B)$ under $H_p$ we need to account for missing information, namely the matching of marked points in $A$ and $B$. To handle this, we first partition $M$ into four parts
\begin{align*} 
&M_{11}=\{m\in M:I_A(m)=1,I_B(m)=1\},\quad
 &M_{10}=\{m\in M:I_A(m)=1,\,I_B(m)=0\},\\
 &M_{01}=\{m\in M:I_A(m)=0,\,I_A(m)=1\},\quad
 &M_{00}=\{m\in M:I_A(m)=0,I_B(m)=0\},
\end{align*}
which are independent and disjoint marked Poisson point processes, all with mark density $g$, and with intensity functions for the locations is
\begin{align*}
&\rho_{11}(r)=\rho(r)\delta_A(r)\delta_B(r),\quad
&\rho_{10}(r)=\rho(r)\delta_A(r)\{1-\delta_B(r)\},\\
&\rho_{01}(r)=\rho(r)\{1-\delta_A(r)\}\delta_B(r),\quad
&\rho_{00}(r)=\rho(r)\{1-\delta_A(r)\}\{1-\delta_B(r)\},
\end{align*}
respectively, see \citet[p.23]{Moller}. Note that $M_{A1}=M_{11}\cup M_{10}$ and $M_{B1}=M_{11}\cup M_{01}$, so $M_{00}$ will play no role in the sequel.  This partitioning is illustrated in \cref{fig:partition}. 

Applying steps A2--A3 to $M_{10}$ yields $M_{103}\sim \MPPP{(\rho_{103},g_{A3})}$, where
\begin{equation}
\rho_{103}(r)=f_A\ast\rho_{10}\{(r-\tau_A)/\psi_A\}/|\psi_A|^2.
\label{eq:rho103}
\end{equation}
Similarly, applying steps B2--B3 to $M_{01}$ yields $M_{013}\sim\MPPP{(\rho_{013},g_{B3})}$ with 
\begin{equation}
\rho_{013}(r)=f_B\ast\rho_{01}\{(r-\tau_B)/\psi_B\}/|\psi_B|^2.
\label{eq:rho013}
\end{equation}
Finally, for each $m\in M_{11}$ we apply steps A2--A3 to yield a marked point $a(m)$, and separately steps B2--B3 to yield a marked point $b(m)$.  The set of paired marked points
\[ M_{113}=\{(a(m), b(m)) : m\in M_{11}\}\]
 forms an $\MPPP$ with paired points in $\mathbb C\times\mathbb C$ and corresponding marks in $(\mathbb S^1\times\{-1,0,1\})^2$. These points have intensity function
\begin{equation}
\rho_{113}(r_a,r_b)= \int_{\mathbb C}\rho_{11}(r)f_A\{(r_a-\tau_A)/\psi_A-r\} f_B\{(r_b-\tau_B)/\psi_B-r\}/|\psi_A\psi_B|^2\,\dd r.
\label{eq:matchdensity}
\end{equation}
The marks are \iid\, with density
\begin{equation}
g_{113}(s_a,t_a,s_b,t_b)=\sum_{u\in\{-1,1\}}d_A(ut_a)d_B(ut_b)\int_{\mathbb S^1}g(s,u)h_A\left(\frac{s_a\conj{s\psi_A}}{|\psi_A|}\right)h_B\left(\frac{s_b\conj{s\psi_B}}{|\psi_B|}\right) \,\dd \nu(s)
\label{eq:convolution:marks}
\end{equation}
\wrt $\mu\times\mu$, and they are independent of the points.

The distribution of $M_{113}$ is dominated by $\zeta_2=\MPPP{(\varphi_2,1/9)}$, the $\MPPP$ whose points form a Poisson point process on $\mathbb C\times\mathbb C$ with intensity function $\varphi_2$ and whose marks are independently uniformly distributed on $(\mathbb{S}^1\times\{-1,0,1\})^2$ and independent of the points. From \cref{eq:matchdensity} we have
\[\int_{\mathbb C\times\mathbb C} \rho_{113}(r_a,r_b) \,\dd r_a \dd r_b= \int_{\mathbb C}\rho_{11}(r)\,\dd r,\]
and hence the density of $M_{113}$ \wrt $\zeta_2$ is
\[\pr(M_{113}\cd \Theta,H_p)=c_2(M_{113})\exp\left\{-\int_{\mathbb C}\rho_{11}(r)\,\dd r\right\}\prod_{(a,b)\in M_{113}}\quad\rho_{113}(r_a,r_b)g_{113}(s_a,t_a,s_b,t_b),\]
where 
\[c_2(M_{113})=9^{|M_{113}|}\exp(1)\quad\prod_{(a,b)\in M_{113}}\{\varphi(r_a)\varphi(r_b)\}^{-1}.\]
Observing that $c(M_{103})c(M_{013})c_2(M_{113})=\exp(1)c(A)c(B)$, the density for $(M_{103},M_{013},M_{113})$ \wrt $\zeta\times\zeta\times\zeta_2$ is 
\begin{multline}
\pr(M_{103},M_{013},M_{113}\cd\Theta,H_p) = c(A)c(B)\exp\left[1-\int_{\mathbb C}\rho(r)\left\{\delta_A(r)+\delta_B(r)-\delta_A(r)\delta_B(r)\right\} \,\dd r\right]\\
\begin{aligned}
&\times\left\{\prod_{a\in M_{103}}\rho_{103}(r_a)g_{A3}(s_a,t_a)\right\} \left\{\prod_{b\in M_{013}}\rho_{013}(r_b)g_{B3}(s_b,t_b)\right\}\\
&\times\left\{\prod_{(a,b)\in M_{113}}\rho_{113}(r_a,r_b)g_{113}(s_a,t_a,s_b,t_b)\right\}.
\end{aligned}
\label{eq:PointsetsMDensity}
\end{multline}

The three marked point processes $(M_{103},M_{013},M_{113})$ can be identified with a labelled bipartite graph $(A,B,\xi)$ of maximum degree one with partitioned vertex set $(A,B)$ and edge set $\xi$.  Specifically, we have the transformation
\[
A=M_{103}\cup \Pi_A(M_{113}),\quad B=M_{013}\cup \Pi_B(M_{113}),\quad \xi= \{\edge{a}{b} : (a,b)\in M_{113}\},\]
where we use the notation $\edge{a}{b}$ for elements of $\xi$, which consist of  edges between marked points, whereas the elements of $(a,b)\in M_{113}$ are the marked points themselves. Furthermore, we have the inverse transformation
\[M_{103} = A\setminus \Pi_A(\xi),\quad M_{013} = B\setminus \Pi_B(\xi),\quad M_{113} = \{(a,b) : \edge{a}{b}\in \xi\},\]
where $\Pi_A$ projects to a marked point set on $\mathbb{M}$ via
\[
\Pi_A(M_{113})=\{a : (a,b)\in M_{113}\mbox{ for some }b\in\mathbb{M}\}.\]
We slightly abuse notation by also writing
\[\Pi_A(\xi)=\{a : \edge{a}{b}\in\xi\mbox{ for some }b\in\mathbb{M}\}.\]
The projector $\Pi_B$ is defined analogously. 

We let $\Xi(A,B)$ denote the space of all possible values for $\xi$, i.e.\ all possible edge sets for the vertex sets $A$ and $B$.  The cardinality of $\Xi(A,B)$ is
\begin{equation*}
|\Xi(A,B)|=
\sum_{n_\xi=0}^{\min(n_A,n_B)}
\frac{n_A!}{n_\xi!(n_A-n_\xi)!}\frac{n_B!}{n_\xi!(n_B-n_\xi)!} n_\xi!,
\end{equation*}
where $n_A,n_B$, and $n_\xi$ denote the cardinality of $A,B$, and $M_{113}$, respectively.  This reflects choosing $n_\xi$ points each from $A$ and $B$ to be matched and considering all $n_\xi!$ edge sets between those points.

Let $\pr(A,B, \xi \cd \Theta, H_p)$ denote the density of $(A,B,\xi)$ \wrt $\tilde\zeta$, where for fixed $(A,B)$,  $\tilde\zeta$ is the counting measure on $\Xi(A,B)$, i.e.\ it holds for $C\subseteq \Xi(A,B)$ that
\begin{equation*}
\dd\tilde \zeta(A,B,C)=|C|\,\dd\zeta(A)\dd\zeta(B).
\end{equation*}
Note that $\sum_{\xi\in\Xi(A,B)}\dd\tilde\zeta(A,B,\xi)=\dd\zeta(A)\dd\zeta(B)$, and thus the marginal density $\pr(A,B\cd\Theta,H_p)$ of the observed points \wrt $\zeta\times \zeta$ is
\begin{equation}
\pr(A,B\cd\Theta,H_p) = \sum_{\xi\in\Xi(A,B)} \pr(A,B,\xi\cd\Theta,H_p).
\label{eq:xisize}
\end{equation}

Now let $\lambda$ denote the distribution of $(A,B, \xi)$ induced by $\zeta\times\zeta\times\zeta_2$, i.e.\ $\lambda$ is the measure $\zeta\times\zeta\times\zeta_2$ transformed by the bijection $(M_{103} ,M_{013},M_{113}) \to (A,B, \xi)$.  Using the expansion for the Poisson process measure \citep[proposition 3.1]{Moller}, a long but straightforward calculation shows that 
$\dd\lambda(A,B,\xi)/\dd \tilde\zeta= \exp(-1)$,
whence
\begin{equation}
\label{eq:expminus1factor}
\pr(A,B, \xi \cd \Theta, H_p)=\exp(-1)\pr(M_{103},M_{013},M_{113}\cd\Theta,H_p).
\end{equation}
\section{Parametric models} 
\label{sect:parametric}

\subsection{Model specification} 

To complete the specification of our basic point process model we need to specify parametric models for the basic elements $(\rho,g,\delta_A,\delta_B,f_A,f_B,h_A,h_B,d_A,d_B)$ introduced in  \cref{sect:generic} that define our marked Poisson point processes and the corresponding likelihood ratios. Clearly there are many possibilities. Below we specify a simple choice to be used in the present paper with the purpose of illustrating and investigating the methodology. We shall return to the potential for improving this choice later. \cite{mythesis} provides a more detailed discussion of the issue.

We assume the intensity $\rho$ and mark density $g$ of $M$ are
\[\rho(r)=\rho_0\varphi(r;\tau_0,\sigma_0^2), \quad g(s,t)=|t|\sqrt{\chi^{|t|+t}(1-\chi)^{|t|-t}},\]
where $\rho_0>0$ and $\chi\in (0,1)$ is the probability that a minutia is a bifurcation.  Note that $g(s,1)=\chi$, $g(s,0)=0$, and $g(s,-1)=1-\chi$.  Without loss of generality, we assume that $\tau_0=0$, since this parameter can be absorbed into $\tau_A$ and $\tau_B$, cf.\ \cref{eq:mappedPointSet}.  Similarly, we assume that $\sigma_0=1$, since this parameter can be absorbed into $\psi_A$ and $\psi_B$.  Due to the latent mark distribution $g(s,t)$ being uniform over $s$, we have 
\[
g_{A1}(s,t)=g(s,t),\quad g_{A2}(s,t)=g_{A3}(s,t)=d_A(t)\chi+d_A(-t)(1-\chi),
\]
and similarly for $B$.

\emph{Thinning.} We assume the selection probabilities are constant with $\delta_A(r)=\delta_A\in (0,1)$ and $\delta_B(r)=\delta_B\in (0,1)$ so that the intensities after thinning become
\[\rho_{A1}(r)=\rho_0\delta_A\varphi(r), \quad \rho_{B1}(r) = \rho_0 \delta_B \varphi(r).\]

\emph{Displacement.}  We assume the error distributions of the minutia locations and types are
\[f_A(r)=f_B(r)=\varphi(r;0,\omega^2),\quad d_A(c)=d_B(c)=I(c=1)\varepsilon+I(c=0)(1-\varepsilon)\]
for some $\varepsilon\in(0,1)$, where $I$ is the indicator function. 
Thus we assume that there are no type misclassifications, though we allow types to be unobserved.  These error functions imply
\begin{gather*}
\rho_{A2}(r)=\rho_0\delta_A\varphi(r;0,1+\omega^2),\quad\rho_{B2}(r)=\rho_0\delta_B\varphi(r;0,1+\omega^2),\\
 g_{A2}(s,t)=g_{B2}(s,t)=(1-|t|)\varepsilon+|t|(1-\varepsilon)g(s,t).
\end{gather*}
The error distributions of the orientations $h_A=h_B=h$ are root von Mises distributions $\rvM{1}{\kappa}$ as defined in \cref{sect:basicdist}.

\emph{Mapping.} After mapping we have
\begin{gather*}
\rho_{A3}(r)= \rho_0\delta_A\varphi\{r;\tau_A,(1+\omega^{2})|\psi_A|^2\},\quad\rho_{B3}(r)= \rho_0\delta_B\varphi\{r;\tau_B,(1+\omega^{2})|\psi_B|^2\},\\
g_{A3}(s,t)=g_{B3}(s,t)=(1-|t|)\varepsilon+|t|(1-\varepsilon)g(s,t).
\end{gather*}
We let $\psi=\psi_A\conj{\psi_B}/(|\psi_A||\psi_B|)$; 
then $\psi$ specifies the relative rotation of $A$ \wrt $B$.
For simplicity we assume in the following that the minutia configurations are represented on the same scale so that $|\psi_A|=|\psi_B|$. further let  $\sigma^{2}=(1+\omega^{2})|\psi_A|^2=(1+\omega^{2})|\psi_B|^2$. 

\subsection{Density under the defence hypothesis}

For the defence likelihood \cref{eq:likelihood-D} we have
\begin{multline}
\pr(A,B\cd \Theta,H_d)=\tilde c(A)\tilde c(B)\exp\left\{-\rho_0(\delta_A+\delta_B)\right\} \rho_0^{n_A+n_B}\delta_A^{n_A}\delta_B^{n_B}\\
\times \chi^{n_A^{(1)}+n_B^{(1)}}(1-\chi)^{n_A^{(-1)}+n_B^{(-1)}}\left\{\prod_{a\in A}\varphi(r_a;\tau_A,\sigma^2)\right\} \left\{\prod_{b\in B}\varphi(r_b;\tau_B,\sigma^2)\right\},
\label{eq:likelihoodHdmarginalized}
\end{multline}
where $n_A^{(t)}=\sum_{a\in A} I(t_a=t)$ for each $t\in\mathbb \{-1,0,1\}$, $\tilde c(A) = c(A)\varepsilon^{n_A^{(0)}}(1-\varepsilon)^{n_A^{(-1)}+n_A^{(1)}}$,
and similarly for $n_B^{(t)}$ and $\tilde c(B)$.  

\subsection{Density under the prosecution hypothesis}
The transformed intensities \cref{eq:rho103}, \cref{eq:rho013}, and \cref{eq:matchdensity} become
\begin{gather}
\rho_{103}(r_a)=\rho_0 \delta_A(1-\delta_B)\varphi(r_a;\tau_A,\sigma^2),\quad
\rho_{013}(r_b)=\rho_0(1-\delta_A)\delta_B\varphi(r_b;\tau_B,\sigma^2),\nonumber\\
\rho_{113}(r_a,r_b)=\rho_0\delta_A\delta_B\varphi_2(r_a,r_b;\tau_A,\tau_B, \Sigma_{AB}),\quad
\Sigma_{AB}=\sigma^2
	\begin{pmatrix}1&\psi/(1+\omega^2)
	\\\conj{\psi}/(1+\omega^2)&1\end{pmatrix}.
	\label{eq:SigmaAB}
\end{gather}
The mark density \cref{eq:convolution:marks} becomes 
\[g_{113}(s_a,t_a,s_b,t_b)=
g_{A2}(s_a,t_a)g_{B2}(s_b,t_b)T(t_a,t_b)\exp\{\kappa\Re(s_a\conj{s_b\psi})\}/I_0(\kappa),\]
where
\begin{equation}
T(t_a,t_b)=(1+t_at_b)\left\{2^4\chi^{|t_a|+t_a+|t_b|+t_b}(1-\chi)^{|t_a|-t_a+|t_b|-t_b}\right\}^{-t_at_b/4}.
\label{eq:typefunction}
\end{equation}
Note that $T(t_a,t_b)=1$ if $t_at_b=0$, $T(t_a,t_b)=0$ if $t_at_b=-1$, $T(1,1)=1/\chi$, and $T(-1,-1)=1/(1-\chi)$. Combining these basic elements with \cref{eq:PointsetsMDensity} and \cref{eq:expminus1factor}, we obtain
\begin{multline}
\pr(A,B,\xi\cd \Theta,H_p)=\tilde c(A)\tilde c(B)\exp\left\{-\rho_0\left(\delta_A+\delta_B-\delta_A\delta_B\right)\right\}\rho_0^{n_A+n_B-n_\xi} \\ 
\begin{aligned}
&\times\chi^{n_A^{(1)}+n_B^{(1)}-n_\xi^{(1)}} (1-\chi)^{n_A^{(-1)}+n_B^{(-1)}-n_\xi^{(-1)}}\delta_A^{n_A}\delta_B^{n_B}(1-\delta_A)^{n_B-n_\xi}(1-\delta_B)^{n_A-n_\xi}\\
&\times\left\{\prod_{a\in A\setminus \Pi_A(\xi)}\varphi(r_a;\tau_A,\sigma^2)\right\} \left\{\prod_{b\in B\setminus \Pi_B(\xi)}\varphi(r_b;\tau_B,\sigma^2)\right\}\\
&\times\left[\prod_{\edge{a}{b}\in\xi} \varphi_2(r_a,r_b;\tau_A,\tau_B, \Sigma_{AB})\frac{\exp\{\kappa\Re(s_a\conj{s_b\psi})\}}{I_0(\kappa)}\right],
\end{aligned}
\label{eq:HpDensityUnexpanded}
\end{multline} 
where $n_\xi^{(t)}=\sum_{\edge{a}{b}\in\xi} I(t_a=t_b=t)$.

\subsection{Variability of parameters}
\label{sect:priors}

The densities in the parametric models specified above depend on 
\begin{equation*}
\Theta=(\rho_0,\chi,\varepsilon,\delta_A,\delta_B,\tau_A,\tau_B,\sigma,\psi, \omega,\kappa),
\end{equation*}
where $\rho_0>0$, $\chi,\varepsilon,\delta_A,\delta_B\in (0,1)$, $\tau_A,\tau_B\in \mathbb C$, $\sigma>0,\psi\in \mathbb S^1,\omega>0$, and $\kappa >0$ are variation independent parameters. As $\tau_A$ and $\tau_B$ are complex numbers there are thirteen real parameters in total.   Of these, $\rho_0$ and $\chi$  relate to the latent minutiae and are common to all fingerprints and fingermarks under consideration. We shall assume the same for $\varepsilon$, $\omega$, and $\kappa$.  The parameters $\rho_0$, $\chi$, $\omega$, and $\kappa$ will be replaced by point estimates and hence treated as being known; we suppress the dependence on these parameters in the following.  Similarly $\varepsilon$ is considered fixed; it only enters via the factors $\tilde c(A)$ and $\tilde c(B)$ which are common to both hypotheses and hence these cancel in the likelihood ratio so $\varepsilon$ can be ignored. This would also be true if we had separate observation probabilities $\varepsilon_A$ and $\varepsilon_B$ for the prints and marks. The remaining parameters
\[\theta=(\delta_A,\delta_B,\tau_A,\tau_B,\sigma,\psi)\]
 vary from one fingerprint or fingermark to the next, according to suitable prior distributions to be specified below. In this way, our approach takes inspiration both from empirical Bayes methods and random effect models.

We follow \citet{compatiblepriors} and ensure that we use compatible prior distributions for the competing models $H_d$ and $H_p$.  Our compatibility condition is that the marginal distributions agree, which leads to the constraint
\begin{equation*}
\int \pr(A\cd\theta)\{\pr(\theta\cd H_p)-\pr(\theta\cd H_d)\}\,\dd\theta
\end{equation*}
for arbitrary values of $A$. For the parametric model described in \cref{sect:parametric}, the constraint becomes
\begin{align*}
\int &\{\pr(\delta_A,\tau_A,\sigma\cd H_p)-\pr(\delta_A,\tau_A,\sigma\cd H_d)\}\\
&\times\exp(-\rho_0\delta_A)\left\{\frac{\delta_A}{\sigma^2}\exp\left(-\frac{|\tau_A|^2-2|\tau_Ar_1|+|r_2|^2}{\sigma^2}\right)\right\}^{n_A}\,\dd(\delta_A,\tau_A,\sigma)=0
\end{align*}
for all $r_1,r_2\in\mathbb C$, and all non-negative integers $n_A$.  The fundamental lemma of the calculus of variations then implies $\pr(\delta_A,\tau_A,\sigma\cd H_p)=\pr(\delta_A,\tau_A,\sigma\cd H_d)$ almost everywhere.  Thus $\delta_A,\delta_B,\tau_A,\tau_B$, and $\sigma$ must have common priors under $H_d$ and $H_p$.  The remaining parameter $\psi$ does not enter under $H_d$ and is thus unconstrained by this consideration.

For our likelihood $\pr(A,B\cd H_p)$ to be invariant under scale transformations, we must require that
\[\pr(A,B,\xi\cd\theta,H_p)\pr(\lambda\tau_A,\lambda\tau_B,\lambda\sigma,\psi)\,\dd(\lambda\tau_A,\lambda\tau_B,\lambda\sigma,\psi)\]
to be independent of the value of $\lambda>0$.   Thus, for the likelihood to be invariant under translation and rotation as well, by \cref{eq:HpDensityUnexpanded} the prior density must be of the form 
\[
\pr(\tau_A,\tau_B,\sigma,\psi\cd H_p)=\sigma^{-5}.
\]
A similar argument shows that $\pr(\tau_A,\tau_B,\sigma,\psi\cd H_d)=\sigma^{-5}$.  This prior density is improper, i.e.\ not integrable over the entire parameter domain. Normally such a prior may result in a meaningless likelihood ratio. However, in our case the improper prior is common to both models $H_d$ and $H_p$ under consideration and the marginal likelihood ratio is equal to the limit of likelihood ratios determined by integrals over the same large box in numerator and denominator.

Under both $H_p$ and $H_d$, we also assume the following.  The fingerprint selection probability $\delta_A$ has a conjugate beta distribution with parameters $(\alpha_\delta, \beta_\delta)$. Assuming that we have a database that is representative for  minutiae in a fingerprint,  these parameters can be estimated reliably. The fingermark selection probability $\delta_B$ has a uniform distribution on $(0,1)$, as it will refer to a fingermark that is not taken from a well-defined population of marks.  Finally we assume that $\delta_A$, $\delta_B,\tau_A,\tau_B,\sigma,$ and $\psi$ are mutually independent.

Thus the joint prior density of the varying parameters is the same under both $H_p$ and $H_d$, and equal to
\begin{equation}
\pr(\theta)=\pr(\theta\cd H_d)=\pr(\theta\cd H_p)=\frac{\Gamma(\alpha_\delta+\beta_\delta)}{\Gamma(\alpha_\delta)\Gamma(\beta_\delta)}\delta_A^{\alpha_\delta-1}(1-\delta_A)^{\beta_\delta-1}\sigma^{-5},
\label{eq:prior}
\end{equation}
where $\Gamma(\cdot)$ is the Gamma function.  We have suppressed the dependence of $\pr(\theta)$ on the hyperparameters $\alpha_\delta$ and $\beta_\delta$. 

Our final model contains the unknown parameters $\rho_0,\chi,\omega,\kappa,\alpha_\delta,\beta_\delta$.  In the developments below we shall consider these parameters as fixed and equal to values estimated from a database of fingerprints and fingermarks as described further in \cref{sect:hyperestimation} below.

\section{Calculating the likelihood ratio}
\label{sect:lrcalc}

\subsection{Defining the likelihood ratio} 

We can in principle obtain our desired likelihood ratio \cref{eq:LR} by summing \cref{eq:HpDensityUnexpanded} over $\xi$, taking its expectation, and dividing by the expectation of \cref{eq:likelihoodHdmarginalized}, where the expectations are with respect to $\theta$.  However, under $H_p$ the number of terms in the sum \cref{eq:xisize} is too large to compute by brute force.  For example, for $n_A=n_B=100$, $|\Xi(A,B)|$ is  approximately equal to  $10^{165}$.  We therefore proceed under $H_p$ by approximating the expectations and the sum using a Monte Carlo sampler to be further discussed below. 

Though some  may prefer to call $\LR$ a Bayes factor, integrated likelihood ratio, or marginal likelihood ratio, we use the term likelihood ratio to conform with standard terminology in forensic science.
\subsection{Integrating the density under $H_d$}
\label{sect:integrateunderHd}

Under $H_d$ we can analytically integrate $\pr(A,B\cd\theta,H_d)\pr(\theta)$ over $\theta$ as follows.  First,
\[\int_{\mathbb C^2} \prod_{a\in A}\varphi(r_a;\tau_A,\sigma^2)\,\dd\tau_A=\frac{\pi^{1-n_A}}{n_A}\sigma^{2(1-n_A)}\exp\left(-{S_A}/{\sigma^2}\right),\]
where $S_A=\sum_{a\in A}\|r_a-r_{A\bullet}\|^2$ is the sum of squared deviations from the average $r_{A\bullet}=n_A^{-1}\sum_{a\in A}r_a$; the integral over $\tau_B$ is analogous.
Second, we can integrate over $\delta_A$ using
\[\int_0^1 e^{-\rho_0\delta_A} \delta_A^{\alpha_\delta+n_A-1} (1-\delta_A)^{\beta_\delta-1}\,\dd \delta_A=e^{-\rho_0}\frac{\Gamma(\alpha_\delta+n_A)\Gamma(\beta_\delta)}{\Gamma(\alpha_\delta+\beta_\delta+n_A)} {}_1F_1(\beta_\delta,\alpha_\delta+\beta_\delta+n_A,\rho_0),\]
where ${}_1F_1$ is the confluent hypergeometric function \citep[chapter 13]{specialfcnshandbook}.  Third, for $\delta_B$, we have
\[\int_0^1 e^{-\rho_0\delta_B} \delta_B^{n_B} \,\dd \delta_B=e^{-\rho_0}\frac{1}{n_B+1}{}_1F_1(1,n_B+2,\rho_0).\]
Fourth, the integral over $\sigma$ is proportional to a gamma density for $\sigma^{-2}$:
\[\int_{0}^\infty\sigma^{-2n_A-2n_B-1}\exp\left\{-(S_A+S_B)/\sigma^2\right\}\,\dd\sigma=\Gamma(n_A+n_B)\left(S_A+S_B\right)^{-n_A-n_B}/2.\]
Combining these integrals with \cref{eq:xisize} and \cref{eq:prior}, the marginal likelihood under $H_d$ is
\begin{multline*}
\pr(A,B\cd H_d)=\tilde c(A)\tilde c(B)e^{-2\rho_0}\pi^{2}\chi^{n_A^{(1)}+n_B^{(1)}}(1-\chi)^{n_A^{(-1)}+n_B^{(-1)}}\left\{\frac{\rho_0}{\pi(S_A+S_B)}\right\}^{n_A+n_B} \\
\begin{aligned}
&\times\frac{\Gamma(\alpha_\delta+\beta_\delta)\Gamma(\alpha_\delta+n_A)\Gamma(n_A+n_B)}{2\Gamma(\alpha_\delta)\Gamma(\alpha_\delta+\beta_\delta+n_A)n_An_B(n_B+1)}
{}_1F_1(\beta_\delta,\alpha_\delta+\beta_\delta+n_A,\rho_0) {}_1F_1(1,n_B+2,\rho_0).
\end{aligned}
\label{eq:HdLikelihoodMarginalized}
\end{multline*}

\subsection{Approximating the likelihood under $H_p$}
\label{sect:ApproxLikelihoodHp}

We are interested in calculating the likelihood ratio $\LR={\pr(A,B\cd H_p)}/{\pr(A,B\cd H_d)}$, cf.\ \cref{sect:lr}, for assessing the  strength of the evidence for $H_p$. We cannot analytically obtain $\pr(A,B\cd H_p)$ because the required sums and integrals are intractable.  Instead we approximate the likelihood ratio using a Markov chain Monte Carlo procedure. There are a variety of possible methods but we have chosen Chib's method \citep{chib95,chib01}.  Other possibilities were investigated in \cite{mythesis}, who found Chib's method to be superior for our specific purpose. Chib's method uses  the simple relation
\[
\pr(A,B\cd H_p) = \frac{\pr(A,B,\theta\fixedaccent,\xi\fixedaccent\cd H_p)}{\pr(\theta\fixedaccent,\xi\fixedaccent\cd A,B,H_p)},
\]
which holds for any fixed values $\theta\fixedaccent$ of $\theta$ and $\xi\fixedaccent$ of $\xi$.  The numerator is simply the product of \cref{eq:HpDensityUnexpanded} and \cref{eq:prior}.  Thus we can approximate $\pr(A,B\cd H_p)$ by approximating the denominator, which can be rewritten as 
\begin{multline*}
\pr(\theta\fixedaccent,\xi\fixedaccent\cd A,B,H_p)=\pr(\delta_A\fixedaccent\cd A,B,H_p)\times \pr(\delta_B\fixedaccent\cd \delta_A\fixedaccent,A,B,H_p)\times \pr(\tau_A\fixedaccent,\tau_B\fixedaccent\cd \delta_A\fixedaccent,\delta_B\fixedaccent,A,B,H_p)\\
\begin{aligned}
&\times \pr(\sigma\fixedaccent\cd \delta_A\fixedaccent,\delta_B\fixedaccent,\tau_A\fixedaccent,\tau_B\fixedaccent,A,B,H_p)\times \pr(\psi\fixedaccent\cd \delta_A\fixedaccent,\delta_B\fixedaccent,\tau_A\fixedaccent,\tau_B\fixedaccent,\sigma\fixedaccent,A,B,H_p)\\
&\times \pr(\xi\fixedaccent\cd \delta_A\fixedaccent,\delta_B\fixedaccent,\tau_A\fixedaccent,\tau_B\fixedaccent,\sigma\fixedaccent,\psi\fixedaccent,A,B,H_p).
\end{aligned}
\end{multline*}
Each of the factors on the right-hand side can be approximated with a suitable sample average of the appropriate full conditional posterior density. The accuracy of these approximations increases with the posterior probability of $(\theta\fixedaccent,\xi\fixedaccent)$. Our method of selecting these values and performing the approximations is detailed in the appendix.

For the final term $\pr(\xi\fixedaccent\cd\theta\fixedaccent,A,B,H_p)$, notice the following. Given a matching $\xi\in\Xi(A,B)$ and a $\beta\in B$, let the sub-matching $\xi_{<\beta}\in\Xi(A,B)$ be given by
\[
\xi_{<\beta} = \{\edge{a}{b}\in\xi : b \in B, b<\beta\},
\]
where the inequality is \wrt some arbitrary total ordering on $B$. Given any $m\in\mathbb M$ and $\dummyelement\in\mathbb M\setminus(A\cup B)$, we define $\Pi_{A,m} : \Xi(A,B)\to\mathbb M$ by
\begin{equation}
\Pi_{A,m}(\xi) = 
\begin{cases}
a&\mbox{ if }\edge{a}{m}\in \xi,\\
\dummyelement& \mbox{ otherwise},
\end{cases}
\label{eq:XiAmProjector}
\end{equation}
which is well-defined because $\xi$ is the edge set of a bipartite graph with maximum degree one, and hence each vertex $m$ is incident with at most one edge $\edge{a}{m}\in\xi$. 

With this notation, we can write
\begin{multline}
\pr(\xi\fixedaccent\cd \theta\fixedaccent,A,B,H_p)
=\prod_{\beta\in B} \E
\left\{\pr(\xi\fixedaccent\cd  \xi\fixedaccent_{<\beta},\xi\fixedaccent_{>\beta},\theta\fixedaccent,A,B,H_p)\,\middle|\,\xi\fixedaccent_{\leq\beta},\theta\fixedaccent,A,B,H_p\right\}
\label{eq:xiprobability}
\end{multline}
where the expectation is over the sub-match $\xi\fixedaccent_{>\beta}$. Notice that the possible values of $\xi\fixedaccent\cd  \xi\fixedaccent_{<\beta},\xi\fixedaccent_{>\beta}$ differ only by which minutia is matched to $\beta$.  By ignoring terms independent of the match of $\beta$, we see from \cref{eq:SigmaAB}--\cref{eq:HpDensityUnexpanded} that
\begin{equation}
\pr(\xi\fixedaccent\cd  \xi\fixedaccent_{<\beta},\xi\fixedaccent_{>\beta},\theta\fixedaccent,A,B,H_p)\propto\exp\left[w\left\{\Pi_{A,\beta}(\xi\fixedaccent),\beta\cd\theta\fixedaccent\right\}\right ]I\{\Pi_{A,\beta}(\xi\fixedaccent)\notin\Pi_A(\xi\fixedaccent_{<\beta}\cup\xi\fixedaccent_{>\beta})\}
\label{eq:fullconditionalofxib}
\end{equation}
where $w$ is
\begin{multline}
w(a,b\cd\theta)=I(a\in A)I(b\in B)\left[\Re\left(\kappa s_a\conj{\psi s_b}  + 2\frac{\omega^2+1}{(\omega^2+1)^2-1}\conj{\psi}\frac{r_a-\tau_A}{\sigma}\frac{\conj{r_b-\tau_B}}{\sigma}\right)\right.\\
\left.   - \frac{1}{(\omega^2+1)^2-1}\left(\frac{|r_a-\tau_A|^2}{\sigma^2}+ \frac{|r_b-\tau_B|^2}{\sigma^2}\right)+\log\left\{\frac{T(t_a,t_b)(\omega^2+1)^2}{\rho_0 I_0(\kappa)(1-\delta_A)(1-\delta_B)\omega^2(\omega^2+2)}\right\}\right].
\label{eq:xiproposal}
\end{multline}
 The normalization constant of \cref{eq:fullconditionalofxib} can be obtained by summing over the support, which is $\xi\fixedaccent_{<\beta}\cup\xi\fixedaccent_{>\beta}$ and $\xi\fixedaccent_{<\beta}\cup\xi\fixedaccent_{>\beta}\cup\{\edge{a}{\beta}\}$ for each $a\in A$.

Thus we can evaluate and normalize \cref{eq:fullconditionalofxib}, and therefore we can approximate \cref{eq:xiprobability} by approximating each expectation with a sample average. Further details are given in the appendix.

\subsection{Sampling procedure}
\label{sect:samplingsummary}

We use a Metropolis-within-Gibbs sampler to generate joint samples of $(\theta,\xi)$ from the posterior distribution $\pr(A,B,\xi\cd\theta,H_p)\pr(\theta)$, the product of \cref{eq:HpDensityUnexpanded} and \cref{eq:prior}.  Our method is detailed in the appendix.  Briefly, we alternate between updating $\delta_A$, $\delta_B$, $(\tau_A,\tau_B)$, $\sigma$, $\psi$, and $\xi$.  We use Gibbs updates for everything except $\xi$: for $\delta_A$ and $\delta_B$ this involves a rejection sampler, while the other updates are straightforward.  For $\xi$, \citet{BayesianAlignmen} propose using a Metropolis--Hastings sampler which creates or breaks a single, random matched pair at each iteration.  However, we have developed a different sampler for $\xi$ which considers all matches for a given minutia simultaneously and computes the probability of each match.  Empirically our sampler appears to converge faster than the sampler in \citeauthor{BayesianAlignmen}.
\section{Data analysis}
\label{sect:datanalysis}

\subsection{Datasets}
\label{sect:dataset}

To investigate the feasibility of our model and algorithm for fingerprint analysis we now apply these to real and simulated data examples.

\begin{figure}[htb]
\center
\includegraphics[width=\textwidth]{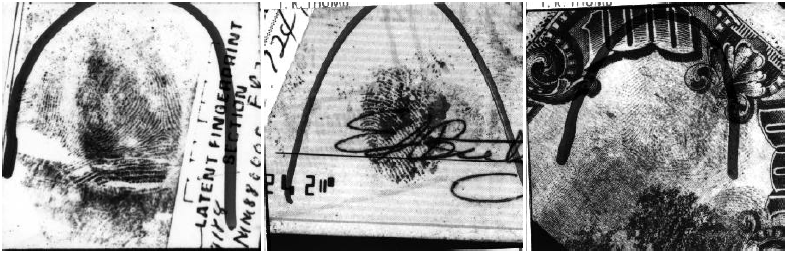} 
\caption{Example fingermarks from \citet{nistdatabase}. From left to right, the fingermark qualities are good, bad, and ugly.}
\label{fig:goodbadugly}
\end{figure}

The real dataset originates from a small database provided by the National Institute for Standards and Technology (NIST) and the Federal Bureau of Investigation (FBI) \citep{nistdatabase}.  This database consists of 258 fingermarks and their corresponding exemplar fingerprints.  The exemplar fingerprints $A$ are all of high quality, and the fingermarks $B$ are of significantly lower quality.  The fingerprint/fingermark pairs are partitioned into three sets based on the quality of the fingermarks: 88 pairs are of relatively good quality, 85 are bad, and 85 are ugly; see \cref{fig:goodbadugly}.    All fingermarks and fingerprint images have their minutiae hand-labelled by expert fingerprint examiners.  This dataset is used for estimation of unknown parameters, for model criticism, and for evaluating the performance of the calculated likelihood ratio.

For reference we also apply our method to data which are simulated from the model using the parameters estimated from the database as described below.  We generated 258 fingerprint/fingermark pairs according to the model described in \cref{sect:parametric} and \cref{sect:priors}.  To ease the comparison with the real database, we also partitioned the simulated data into a good set consists of those 88 pairs with the highest number of fingermark minutiae $n_B$, a bad set containing the next 85 pairs, and an ugly set containing those $85$ pairs with the lowest $n_B$.  By comparing our results on the NIST database to our results on the simulated data we are able to distinguish model inadequacies from algorithm errors or performance issues.

\subsection{Model criticism} The question of model accuracy was investigated in \citet[chapter 7]{mythesis};  it is apparent that some of the model features are oversimplified and the data behaviour deviates from the assumptions. For example, our model assumes the minutia are independently thinned with constant thinning frequency, have independent orientations, and have independent spatial observation errors.  In fact, the thinning, orientations, and location distortions appear to be correlated amongst nearby minutiae.  We abstain from giving the details here and choose  to proceed with the simple model  despite its apparent shortcomings.

\subsection{Parameter estimation}
\label{sect:hyperestimation}

We must find point estimates for the fixed parameters $\alpha_\delta,\beta_\delta,\rho_0,\chi,\omega,$ and $\kappa$.  As our real dataset contains matched fingerprint/fingermark pairs which conform with the prosecution hypothesis, we estimate all parameters under $H_p$.

The estimates are difficult to find without knowing the correct matching $\xi$.  Unfortunately our dataset contains only 258 paired minutia configurations without matching the corresponding minutiae within a configuration; that is, it contains $A_i$ and $B_i$ but not $\xi_i$ for $i=1\ldots258$.  Previous research \citep{mikalyan} attempted to ameliorate this by running an automated matching algorithm on the dataset. However, we found the quality of these matchings to be extremely poor and instead we manually found and recorded what we believe to be the correct minutia matchings $\check\xi$ for each of the 258 fingerprint/fingermark pairs in the dataset \citep{nistdatabase}.  With this matching $\check\xi$ fixed, we proceeded with the parameter estimation. We emphasize that $\check\xi$ is only used for estimation of the unknown parameters of the model and not otherwise for the calculation of likelihood ratios.

We estimate the fixed parameters by maximizing the likelihood function under $H_p$ and based on matching-augmented data $(A_i,B_i,\check\xi_i)$, i.e.\ 
\[
\prod_{i=1}^{258} \left\{ \int \pr(A_i,B_i,\check\xi_i,\theta_i\cd H_p)\,\dd\theta_i\right\} = \prod_{i=1}^{258} \pr(A_i,B_i,\check\xi_i\cd H_p; \alpha_\delta,\beta_\delta,\rho_0,\chi,\omega,\kappa),
\]
where $\pr(A_i,B_i,\check\xi_i,\theta\cd H_p)$ is the product of \cref{eq:HpDensityUnexpanded} and \cref{eq:prior}, and where the fixed parameters have been suppressed on the left-hand side of this equation.    Each integrand on the left-hand side further  factorizes into 
\begin{align*}
\pr(A_i,B_i,\check\xi_i,\theta\cd H_p)=&f_0(A_i,B_i,\check\xi_i)\times f_1(A_i,B_i,\check\xi_i,\delta_A,\delta_B; \alpha_\delta,\beta_\delta,\rho_0)\\
&\times f_2(A_i,B_i,\check\xi_i; \chi) \times f_3(A_i,B_i,\check\xi_i,\tau_A,\tau_B,\sigma,\psi;\omega,\kappa).
\end{align*}
Here $f_0$ is independent of the parameters we are estimating and thus of no importance. Further
\begin{multline*}
f_1(A,B,\check\xi,\delta_A,\delta_B; \alpha_\delta,\beta_\delta,\rho_0)=\exp\left\{-\rho_0\left(\delta_A+\delta_B-\delta_A\delta_B\right)\right\}\frac{\Gamma(\alpha_\delta+\beta_\delta)}{\Gamma(\alpha_\delta)\Gamma(\beta_\delta)}\\
\times\rho_0^{n_A+n_B-n_\xi}\delta_A^{\alpha_\delta+n_A-1}\delta_B^{n_B}(1-\delta_A)^{\beta_\delta+n_B-n_\xi-1}(1-\delta_B)^{n_A-n_\xi},
\end{multline*}
\[f_2(A,B,\check\xi; \chi)=\chi^{n_A^{(1)}+n_B^{(1)}-n_\xi^{(1)}}(1-\chi)^{n_A^{(-1)}+n_B^{(-1)}-n_\xi^{(-1)}},\]
and
\begin{multline*} f_3(A,B,\check\xi,\tau_A,\tau_B,\sigma,\psi;\omega,\kappa)=\sigma^{-2(n_A+n_B)-5}\left\{\frac{(\omega^2+1)^2}{(\omega^2+1)^2-1}\right\}^{n_\xi}I_0(\kappa)^{-n_\xi}\\
\begin{aligned}
&\times\exp\left\{-\left(\sum_{a\in A\setminus \Pi_A(\xi)}\frac{|r_a-\tau_A|^2}{\sigma^2}\right)   -\left(\sum_{b\in B\setminus \Pi_B(\xi)} \frac{|r_b-\tau_B|^2}{\sigma^2} \right)\right\}\\
&\times\exp\left\{\sum_{\edge{a}{b}\in\xi}\Re\left(\kappa s_a\conj{\psi s_b}  + 2\frac{\omega^2+1}{(\omega^2+1)^2-1}\conj{\psi}\frac{r_a-\tau_A}{\sigma}\frac{\conj{r_b-\tau_B}}{\sigma}\right)\right\}\\
 &\times\exp\left\{- \frac{(\omega^2+1)^2}{(\omega^2+1)^2-1}\sum_{\edge{a}{b}\in\xi}\left(\frac{|r_a-\tau_A|^2}{\sigma^2} + \frac{|r_b-\tau_B|^2}{\sigma^2}\right)\right\}.
\end{aligned}
\end{multline*}

Since $(\alpha_\delta,\beta_\delta,\rho_0)$ only enter into $f_1$, the estimates for these parameters are the maximizers of
\[
\prod_{i=1}^{258} \left\{\int f_1(A_i,B_i,\check\xi_i,{\delta_A}_i,{\delta_B}_i; \alpha_\delta,\beta_\delta,\rho_0) \,\dd({\delta_A}_i,{\delta_B}_i)\right\}.
\]
The integral over $\delta_B$ can be obtained analytically as in \cref{sect:integrateunderHd}.  The integral over $\delta_A$ can be found numerically, and the resulting function can also be maximized numerically.  We used the R package {pracma} for the integrals and the standard R function {optim} for the optimization. The resulting estimates are $\hat\alpha_\delta=$14$\cdot$67, $\hat\beta_\delta=$3$\cdot$30, and $\hat\rho_0=$132$\cdot$74.

Similarly, $\chi$ only enters into $f_2$ and can be found by directly maximizing $\sum_{i=1}^{258}\log f_2(A_i,B_i,\check\xi_i; \chi)$, yielding a linear equation for $\chi$ with the solution $\hat\chi=$0$\cdot$38.

We estimate $\omega$ and $\kappa$ by maximizing the third factor in the likelihood function
\[\prod_{i=1}^{258} \left\{\int f_3(A_i,B_i,\check\xi_i,{\tau_A}_i,{\tau_B}_i,\sigma_i,\psi_i;\omega,\kappa) \,\dd({\tau_A}_i,{\tau_B}_i,\sigma_i,\psi_i)\right\}.\]
This function is too complicated to maximize using standard numerical techniques.  We resort to a stochastic expectation-maximization algorithm \citep{StochasticEM} based on the Monte Carlo Markov chain procedure described in the appendix.  We fix $\alpha_\delta,\beta_\delta,\rho_0$, and $\chi$ to their estimated values above.  Starting from some initial values for $\omega$ and $\kappa$, we generate a posterior sample $({\tau_A}_i,{\tau_B}_i,\sigma_i,\psi_i)$ for each fingerprint/fingermark pair $i=1,\ldots,258$.  We then maximize 
\[\prod_{i=1}^{258} f_3(A_i,B_i,\check\xi_i,{\tau_A}_i,{\tau_B}_i,\sigma_i,\psi_i;\omega,\kappa) \]
over $\omega$ and $\kappa$.  The maximizing value for $x=(\omega^2+1)^2/\{(\omega^2+1)^2-1\}$ is a root of the polynomial equation
\[(R_2^2-4R_3^2)x^3+(4R_3^2-2R_1R_2-R_2^2)x^2+(R_1^2+2R_1R_2-R_3^2)x-R_1^2=0,\]
where 
$R_1=\sum_{i=1}^{258} |\check\xi_i|$,
 $R_2=\sum_{i=1}^{258}\left\{\sigma_i^{-2}\sum_{\edge{a}{b}\in\check\xi_i}\left|r_a-{\tau_A}_i|^2 + |r_b-{\tau_B}_i|^2\right)\right\},$
 and 
\[R_3=\sum_{i=1}^{258} \left[\sigma_i^{-2}\sum_{\edge{a}{b}\in\check\xi_i}\Re\left\{(r_a-{\tau_A}_i)\conj{\psi_i(r_b-{\tau_B}_i)}\right\}\right].\]
 This can be solved using the cubic formula.  The maximizing value for $\kappa$ solves
\[R_1\frac{I_1(\kappa)}{I_0(\kappa)}=\sum_{i=1}^{258}\left\{\sum_{\edge{a}{b}\in\check\xi_i}\Re\left(s_a\conj{\psi_i s_b}\right)\right\},\]
where $I_1$ is the modified Bessel function of the first kind and first order.  The ratio $I_1(\kappa)/I_0(\kappa)$ is always between zero and one, so this equation is simple to solve numerically.

We repeat the process of generating  new values of $({\tau_A}_i,{\tau_B}_i,\sigma_i,\psi_i)$ and updating $\omega$ and $\kappa$ until the latter stabilize.  After they stabilize we run 500 more iterations while saving the maximizing values of $\omega$ and $\kappa$.  Our point estimates for $\omega$ and $\kappa$ are the average of these maximizing values, yielding   $\hat{\omega}=$0$\cdot$047 and $\hat\kappa=$35.

\subsection{Results}

The  Monte Carlo Markov chain algorithm was programmed in C\# version 4$\cdot$51.  We chose this language due to its multi-thread support for multiple parallel fingerprint comparisons and advanced data visualization capabilities.  Our algorithm generates approximately $5000$ joint samples of $\theta$ and $\xi$ per thread per second on a 3GHz Intel Xeon processor.

For both simulated and real data we set the initial value of $\xi$  to the empty match.  Within $2000$ iterations the variable traces appeared to be stationary.  We used $5000$ samples for burn-in and generated another $5000$ samples to estimate the likelihood ratio. In our experience this sample size is sufficient to reduce the Monte Carlo error in the log likelihood ratio estimate to less than 0$\cdot$2.

We computed the log likelihood ratio for all possible $258\times258$ fingerprint/fingermark pairs in our simulated dataset. The log likelihood ratios for the $258$ pairs that originate from the same finger are shown in the blue histogram with solid lines at the top of \cref{fig:LRChibsRnd}.  The remaining $258\times257$ log likelihood ratios (the false matches) are shown in the red histogram with dashed lines. The inset receiver-operating characteristic curve describes our discrimination of true matches from false matches based on any chosen cutoff point for the log likelihood ratio. 

The other three histograms subdivide the pairs into the $88\times 258$ pairs where the fingermark is good, the $85\times258$ pairs where the fingermark is bad, and the $85\times258$ pairs where the fingermark is ugly.  We achieve perfect separation for the good and bad fingermarks, and worse separation for ugly fingermarks, reflecting that these have fewer minutiae and thus are less informative.
 

The same type of histograms for our real dataset  are displayed in \cref{fig:LRChibs}.  The discrimination here is not as good as for the simulated data; this could be another indication that our model does not completely describe the variability in real fingermarks.  
\foreach \resultset/\appendToCaption in {ChibsRnd/simulated ,Chibs/NIST-FBI }{
	\begin{figure}[htp]
	\begin{tikzpicture}
	\foreach \dataset/\datasetDescription [count=\c] in {/full dataset,G/good subset,B/bad subset,U/ugly subset}
	{
		\pgfmathsetmacro\yshift{-\c * 4.7}
		\begin{axis}[xmin=-63,xmax=103,ymin=0,ymax=.15,width=\textwidth,height=5cm, xlabel=$\log_{10}$ of likelihood ratio for \datasetDescription, ylabel=Density, scaled y ticks = false, y tick label style={/pgf/number format/fixed},name=\dataset\c,tick pos=left,yshift=\yshift cm] 
		\addplot[const plot,color=blue,fill opacity=0.7,no markers,thick] table {plotData/\resultset/LRtrueHist\dataset.txt};
		\addplot[const plot,color=red,fill opacity=0.7,no markers,thick,dashed] table {plotData/\resultset/LRfalseHist\dataset.txt};
		\end{axis}
		\begin{axis}[at={(\dataset\c.north east)},anchor=north east,xshift=-0.1cm,yshift=-0.1cm,width=4.2cm,xmin=-0.02,xmax=1.02,ymin=-.02,ymax=1.02, tick pos=left] 
		\addplot[const plot, draw=black] table {plotData/\resultset/ROC\dataset.txt};
		\end{axis}
	}
	\end{tikzpicture}
	\caption{Histogram of the log-likelihood ratios for \appendToCaption data.  Log-likelihood ratios corresponding to false matches are dashed and red, and true matches are solid and blue. Inset is a receiver-operating characteristic curve with the rate of false positives (i.e., the type 1 error rate) on the x-axis and the rate of true positives (i.e., one minus the type 2 error rate) on the y-axis.} 
	\label{fig:LR\resultset}
	\end{figure}
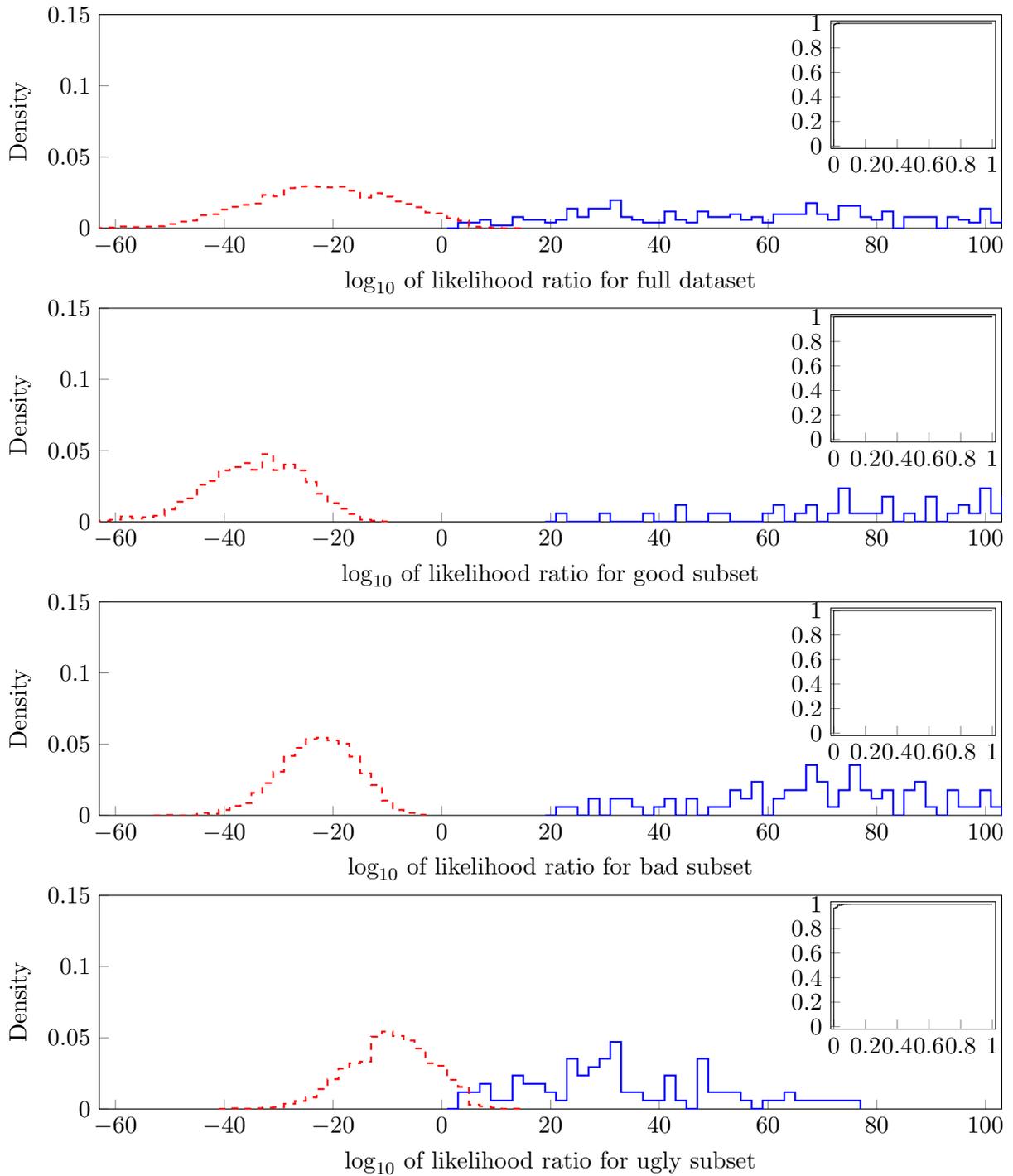	
}

Similarly, the likelihood ratios appear to be slightly more extreme than they should be for the false pairings; for example,  the maximal value of $\log_{10}\LR$ is equal to 12$\cdot$2, which appears too high to occur by chance, and higher than the similar value for simulated data, which is 9$\cdot$6.  
\section{Discussion}
We have described a marked Poisson point process model for paired minutia configurations in fingerprints and fingermarks, and the corresponding matching between these minutia configurations.  We can efficiently sample from the distribution of the unknown matching and parameters in this model using a Markov chain Monte Carlo method.  The resulting sample can be used to compute likelihood ratios for comparing the hypothesis that the two configurations originate from the same finger against the hypothesis that they originate from different fingers.  

The method provides excellent discrimination on simulated data. Using the method on a specific NIST-FBI database  indicate that the model yields good discrimination between these two hypotheses as long as the fingermark is of reasonable quality.  However, some inaccuracies are apparent for the simplistic model discussed in the present paper, in particular concerning the model for selection of observed fingermarks which appears to be non-constant; also, the minutiae do tend to occur along fingerprint ridges which means that orientations of nearby minutiae are not independent, as assumed. 

This may result in the likelihood ratios calculated to be more extreme than what can be justified.  The ratios can still be used as a sensible model based method for discrimination between true and false matches, but they would have to be calibrated against a large real dataset along the lines described in \citet[chapter 9]{mythesis} before they can be interpreted as an accurate measure of the strength of evidence. In any case, we believe the framework can be used to establish a sound and model-based foundation for the analysis of fingerprint evidence.

\section{Acknowledgements}
This research was partially supported by the Danish Council for Independent Research~$\mid$~Natural Sciences and by the Centre for Stochastic Geometry and Advanced Bioimaging.

\appendix
\section*{Appendix}

\subsection*{Overview of the sampling procedures}
Chib's method as discussed in \cref{sect:ApproxLikelihoodHp}, including our choice of $\theta\fixedaccent$ and $\xi\fixedaccent$, is detailed in \cref{alg:chibs}.
\begin{algorithm}[htbp]
\tcc{All samples $S_i$ are generated by holding the starred variables constant while sampling the non-starred variables as described in \cref{alg:gibbs}.}
 $r\gets 1$\;
 Generate sample $S_1$ of size $N$ from $\delta_A,\delta_B,\tau_A,\tau_B,\sigma,\psi,\xi\cd A,B,H_p$\;
 $\delta_A\fixedaccent\gets\hat\E_{S_1}(\delta_A)$\;
 Generate sample $S_2$ of size $N$ from $\delta_B,\tau_A,\tau_B,\sigma,\psi,\xi\cd \delta_A\fixedaccent,A,B,H_p$\;
 $\delta_B\fixedaccent\gets\hat\E_{S_2}(\delta_B)$\;
 $r\gets r\times\hat\E_{S_2}\{\pr(\delta_A\fixedaccent\cd \delta_B,\tau_A,\tau_B,\sigma,\psi,\xi,A,B,H_p)\}$\;
 Generate sample $S_3$ of size $N$ from $\tau_A,\tau_B,\sigma,\psi,\xi\cd \delta_A\fixedaccent,\delta_B\fixedaccent,A,B,H_p$\;
 $(\tau_A\fixedaccent,\tau_B\fixedaccent)\gets\hat\E_{S_3}(\tau_A,\tau_B)$\;
 $r\gets r\times\hat\E_{S_3}\{\pr(\delta_B\fixedaccent\cd \delta_A\fixedaccent,\tau_A,\tau_B,\sigma,\psi,\xi,A,B,H_p)\}$\;
 Generate sample $S_4$ of size $N$ from $\sigma,\psi,\xi\cd \delta_A\fixedaccent,\delta_B\fixedaccent,\tau_A\fixedaccent,\tau_B\fixedaccent,A,B,H_p$\;
 $\sigma\fixedaccent\gets\hat\E_{S_4}(\sigma)$\;
 $r\gets r\times\hat\E_{S_4}\{\pr(\tau_A\fixedaccent,\tau_B\fixedaccent\cd \delta_A\fixedaccent,\delta_B\fixedaccent,\sigma,\psi,\xi,A,B,H_p)\}$\;
 Generate sample $S_5$ of size $N$ from $\psi,\xi\cd \delta_A\fixedaccent,\delta_B\fixedaccent,\tau_A\fixedaccent,\tau_B\fixedaccent,\sigma\fixedaccent,A,B,H_p$\;
 $\psi\fixedaccent\gets\hat\E_{S_5}(\psi)$\;
 $r\gets r\times\hat\E_{S_5}\{\pr(\sigma\fixedaccent\cd \delta_A\fixedaccent,\delta_B\fixedaccent,\tau_A\fixedaccent,\tau_B\fixedaccent,\psi,\xi,A,B,H_p)\}$\;
 Generate sample $S_6$ of size $N$ from $\xi\cd \delta_A\fixedaccent,\delta_B\fixedaccent,\tau_A\fixedaccent,\tau_B\fixedaccent,\sigma\fixedaccent,\psi\fixedaccent,A,B,H_p$\;
 $r\gets r\times\hat\E_{S_6}\{\pr(\psi\fixedaccent\cd \delta_A\fixedaccent,\delta_B\fixedaccent,\tau_A\fixedaccent,\tau_B\fixedaccent,\sigma\fixedaccent,\xi,A,B,H_p)\}$\;
 $\xi\fixedaccent\gets\mathrm{argmax}_\xi\{\pr(\theta\fixedaccent,\xi,A,B,H_p)\}$\tcc*{An efficient method for finding the maximizer over $\xi$ is given in \citet[chapter 3]{mythesis}}
\For{$\beta\in B$}{
 Generate sample $S_{\beta}$ of size $N_\xi$ from $\xi\cd \xi_{\leq\beta}=\xi\fixedaccent_{\leq\beta},\theta\fixedaccent,A,B,H_p$\;
 $r\gets r\times\hat\E_{S_{\beta}}\left\{\pr\left(\xi \cd  \xi_{<\beta},\xi_{>\beta},\theta\fixedaccent,A,B,H_p\right)\right\}$\tcc*{See \cref{eq:fullconditionalofxib}}
}
\KwRet $r$, an estimate of $\pr(A,B\cd H_p)$\;
\caption{Chib's method for approximating $\pr(A,B\cd H_p)$.}
\label{alg:chibs}
\end{algorithm}
The Metropolis-within-Gibbs sampler discussed in \cref{sect:samplingsummary} is described in \cref{alg:gibbs}. 
\begin{algorithm}[htbp]
\KwIn{$\theta^0,\xi^0$ set to some initial value}
\For{$n=1,\ldots,N$}{
 $\delta_A^{n}\gets\sample{\delta_A\cd A,B,\delta_B^{n-1},\tau_A^{n-1},\tau_B^{n-1},\sigma^{n-1},\psi^{n-1},\xi^{n-1}}$\;
 $\delta_B^{n}\gets\sample{\delta_B\cd A,B,\delta_A^{n},\tau_A^{n-1},\tau_B^{n-1},\sigma^{n-1},\psi^{n-1},\xi^{n-1}}$\;
 $(\tau_A^{n},\tau_B^{n})\gets\sample{\tau_A,\tau_B\cd A,B,\delta_A^{n},\delta_B^{n},\sigma^{n-1},\psi^{n-1},\xi^{n-1}}$\;
 $\sigma^{n}\gets\sample{\sigma\cd A,B,\delta_A^{n},\delta_B^{n},\tau_A^n,\tau_B^n,\psi^{n-1},\xi^{n-1}}$\;
 $\psi^n\gets\sample{\psi\cd A,B,\delta_A^{n},\delta_B^{n},\tau_A^n,\tau_B^n,\sigma^{n},\xi^{n-1}}$\;
 $\xi^n\gets\xi^{n-1}$\;
\For{$j=1,\ldots,n_A$}{\tcc{Sample repeatedly to reduce autocorrelation}
 $\xi^n\gets\sample{\xi\cd A,B,\delta_A^{n},\delta_B^{n},\tau_A^n,\tau_B^n,\sigma^{n},\psi^n,\xi^{n}}$\;
}
}
\caption{Metropolis-within-Gibbs sampler for the posterior of $\theta$ and $\xi$.}
\label{alg:gibbs}
\end{algorithm}
To generate our samples, we use \citet{ziggurat} for Gaussian variables, \citet{gammasampler} for Gamma variables, \citet{dagpunar} for truncated Gamma variables, \citet{betasampler} for Beta variables, and \citet{bestfisher} for von Mises variables.  For those variables whose full conditionals are not one of the above type, we give a detailed sampling algorithm below.  All samplers use \citet{xor} as source of pseudo-random integers.

\subsection*{Sampling $\delta_A,\delta_B$}
Define the distribution $D(\alpha,\beta,\lambda)$ to have density
\begin{equation*}
f_D(\delta)\propto \delta^{\alpha-1}(1-\delta)^{\beta-1}e^{-\lambda \delta}
\end{equation*}
for $\delta\in(0,1)$, where $\alpha>0,\beta>0,$ and $\lambda\in\mathbb R$.  If $\lambda=0$ this is a Beta distribution, and if $\beta=1$ it is a Gamma distribution right-truncated at one.  The full conditionals for $\delta_A$ and $\delta_B$ are
\[
\delta_A\sim D(\alpha_\delta+n_A, \beta_\delta+n_B-n_\xi, \rho_0-\rho_0\delta_B),\quad
\delta_B\sim D(n_B, n_A-n_\xi, \rho_0-\rho_0\delta_A).
\]
We describe an algorithm to sample from $D$ in \cref{alg:thinningsampler}.  
\begin{algorithm}[htbp]
\If{$\lambda=0$}{ 
 $\delta\gets\sample{\mathrm{Gamma}(\alpha,\beta)\cd\delta\leq 1}$\;
}\Else{
{$\delta_0\gets \argmax\{ x^{\alpha-1}(1-x)^{\beta-1}e^{-\lambda x} : x\in[0,1]\}$}\;
\If{$\delta_0>0.5$\textbf{ and }$\lambda< (\alpha-1)/(1-\delta_0)$}{
 $\delta\gets 1-\sample{D(\beta,\alpha,-\lambda)}$\;
}\Else{
\Repeat{$U^{1/(\beta-1)}<(1-\delta)\delta_1\exp\{(\delta-\delta_0)/(1-\delta_0)\}$}{
 $\delta\gets\sample{\mathrm{Gamma}\{\alpha,\lambda+(\beta-1)/(1-\delta_0)\}\cd\delta\leq 1}$\;
 $U\gets\sample{\mathrm{Uniform}(0,1)}$\;
}
}
}
\KwRet $\delta$\;
\caption{Rejection sampler for $D(\alpha,\beta,\lambda)$.}
\label{alg:thinningsampler}
\end{algorithm}

Briefly, let $\delta_0$ be the mode of $D$, which can be easily computed by applying the quadratic formula to $\dd\log f_D(\delta)/\dd\delta=0$.  To sample from $D$, first notice that $\log(1-\delta)\approx 1-\delta_0-\delta/(1-\delta_0)$ for small $\delta_0$ ($\delta_0\leq 0.5$ in \cref{alg:thinningsampler}), and hence $(1-\delta)^{\beta-1}\approx C\exp\{-(\beta-1)\delta/(1-\delta_0)\}$ where $C$ is a constant independent of $\delta$.  Plugging this approximation into $f_D(\delta)$ yields a $\mathrm{Gamma}\{\alpha,(\beta-1)/(1-\delta_0)+\lambda\}$ density right-truncated at one.  Thus when $\delta_0\leq 0.5$, we can use rejection sampling with proposals drawn from this distribution.  Similarly, when the mode $\delta_0\geq 0.5$, let $\tilde\delta=1-\delta$ so that $\tilde\delta\sim D(\beta-1,\alpha-1,-\lambda)$ with mode $\tilde\delta_0=1-\delta_0$.  Using the same approximation as before, we can use rejection sampling on $\tilde\delta$ with a $\mathrm{Gamma}(\beta,(\alpha-1)/(1-\tilde\delta_0)-\lambda)$ proposal, right-truncated at one, provided $(\alpha-1)/(1-\tilde\delta_0)-\lambda>0$.  In practice we achieve acceptance rates greater than $0.9$.

\subsection*{Sampling $\tau_A,\tau_B$}
The full conditional is bivariate complex normal with mean $(K_d+n_\xi\Sigma_{AB}^{-1})^{-1}(K_d r_d + n_\xi\Sigma_{AB}^{-1} r_p)$ and inverse variance $K_d+n_\xi\Sigma_{AB}^{-1}$, where
\[
r_{d}=\begin{pmatrix}n_A^{-1}\sum_{a\in A}r_a \\ n_B^{-1}\sum_{b\in B}r_b\end{pmatrix}, \quad
 r_p=\frac{1}{n_\xi}\sum_{\edge{a}{b}\in\xi}\begin{pmatrix}r_a \\ r_b \end{pmatrix}, \quad
K_d = \sigma^{-2}\begin{pmatrix}n_A & 0 \\ 0 & n_B\end{pmatrix}.
\]

\subsection*{Sampling $\sigma$}
We make the change of variables $u=\sigma^{-2}$.  The improper prior $\pr(\sigma)=\sigma^{-5}$ becomes $\pr(u)\propto u$. The full conditional of $u$ is a Gamma distribution with shape parameter $n_A+n_B+2$ and inverse scale parameter
\begin{multline*}
\sum_{a\in A\setminus \Pi_A(\xi)}|r_a-\tau_A|^2+\sum_{b\in B\setminus \Pi_B(\xi)}|r_b-\tau_B|^2\\
+\frac{(\omega^2+1)^2}{(\omega^2+1)^2-1}\sum_{\edge{a}{b}\in\xi}\left[|r_a-\tau_A|^2+|r_b-\tau_B|^2-\frac{2}{\omega^2+1}\Re\left\{(r_a-\tau_A)\conj{(r_b-\tau_B)\psi}\right\}\right].
\end{multline*}

\subsection*{Sampling $\psi$}

The full conditional of $\psi$ is a von Mises distribution with location parameter $\nu_0/|\nu_0|$ and concentration parameter $|\nu_0|$, where
\[\nu_0=\sum_{\edge{a}{b}\in\xi}\kappa s_a\conj{s_b}  + 2\frac{\omega^2+1}{(\omega^2+1)^2-1}\frac{r_a-\tau_A}{\sigma}\frac{\conj{r_b-\tau_B}}{\sigma}.\]

\subsection*{Sampling $\xi$}
\label{sect:xiSampler}

Finally we sample the matching $\xi$.  A possible Metropolis--Hastings sampler for $\xi$ is described in \citet{BayesianAlignmen}.  They propose creating or breaking a single, random matched pair at each iteration.  In contrast, our \cref{alg:xiSampler} considers all matches for a given minutia simultaneously and computes the probability of each match.  

We need one more piece of notation.  In analogy with $\Pi_{A,m}$ in \cref{eq:XiAmProjector}, for each $m\in\mathbb M$, define $\Pi_{B,m} : \Xi(A,B)\to\mathbb M$ by $\Pi_{B,m}(\xi) = b$ if $\edge{m}{b}\in \xi$ for some $b\in B$,  and $\dummyelement$ otherwise.

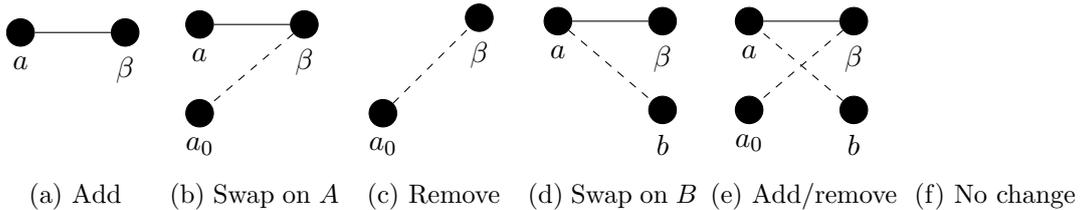
\begin{figure}[thbp]
\centering
    \begin{subfigure}[b]{0.14\textwidth}
        \centering
        \begin{tikzpicture}
			[circ/.style={circle,draw=black,fill=black,minimum size=2pt}]
			\node[circ,label=below:$a$] (a)  {};				
			\node[circ,label=below:${\beta}$,right=of a] (b) {}		edge (a);
				\node[minimum size=2pt,label=below:\phantom{$a_0$},draw=white,fill=white,below=8mm of a] (ap) {};
		\end{tikzpicture}
        \caption{Add}\label{fig:qadd}
    \end{subfigure}
    \begin{subfigure}[b]{0.14\textwidth}
        \centering
        \begin{tikzpicture} 
			[circ/.style={circle,draw=black,fill=black,minimum size=2pt}]
			\node[circ,label=below:$a$] (a)  {};		
			\node[circ,label=below:$a_0$,below=8mm of a] (ap) {}; 
			\node[circ,label=below:$\beta$,right=of a] (b) {}
			edge (a) 
			edge[dashed] (ap);
		\end{tikzpicture}
        \caption{Swap  on $A$}\label{fig:qaddremove}
    \end{subfigure}
    \begin{subfigure}[b]{0.14\textwidth}
        \centering
        \begin{tikzpicture} 
			[circ/.style={circle,draw=black,fill=black,minimum size=2pt}]
			\node[circ,label=below:${\beta}$] (b)  {};				
			\node[circ,label=below:${a_0}$,below left=of b] (a) {}
			edge[dashed] (b);
			\node[below=8mm of b] {};
		\end{tikzpicture}
        \caption{Remove}\label{fig:qremove}
    \end{subfigure}
    \begin{subfigure}[b]{0.14\textwidth}
        \centering
        \begin{tikzpicture}
			[circ/.style={circle,draw=black,fill=black,minimum size=2pt}]
			\node[circ,label=below:${\beta}$] (b) {};		
			\node[circ,below=8mm of b,label=below:$b$] (bp) {} ;
			\node[circ,label=below:$a$,left=of b] (a)  {}
			edge[dashed] (bp)
			edge (b);
							\node[minimum size=2pt,label=below:\phantom{$a_0$},draw=white,fill=white,below=8mm of a] (ap) {};
		\end{tikzpicture}
        \caption{Swap on $B$}\label{fig:qswap}
    \end{subfigure} 
    \begin{subfigure}[b]{0.16\textwidth}
        \centering
        \begin{tikzpicture}
			[circ/.style={circle,draw=black,fill=black,minimum size=2pt}]
			\node[circ,label=below:$a$] (a)  {};	
			\node[circ,label=below:$a_0$,below=8mm of a] (ap) {};			
			\node[circ,right=of ap,label=below:$b$] (b) {}
			edge[dashed] (a); 
			\node[circ,label=below:$\beta$,right=of a] (bp) {}
			edge (a)
			edge[dashed] (ap);
		\end{tikzpicture} 
        \caption{Add/remove}\label{fig:qswapremove}
    \end{subfigure}
		\begin{subfigure}[b]{0.14\textwidth}
		  \centering
        \begin{tikzpicture}[circ/.style={circle,draw=white,fill=white,minimum size=2pt}]
			\node[circ,label=below:\phantom{${\beta}$}] (b)  {};				
			\node[circ,label=below:\phantom{${a_0}$},below left=of b] (a) {};
			\node[below=8mm of b] {};
		\end{tikzpicture}
		\caption{No change}
		\end{subfigure}
\caption{Illustration of which states for $\xi'$ are accessible from a given state $\xi$.  Dashed edges are removed matches, solid edges are added matches.  Edges that are common to both $\xi'$ and $\xi$ are not shown.  We write $a$ for $\Pi_{A,\beta}(\xi')$, $a_0$ for $\Pi_{A,\beta}(\xi)$ and $b$ for $\Pi_{B,a}(\xi)$, assuming that none of these are equal to $\dummyelement$.  Hence $\edge{a_0}{\beta}\in\xi$, $\edge{a}{b}\in\xi$, and $\edge{a}{\beta}\in\xi'$.}
\label{fig:SwapsAllowed}
\end{figure}

We will sample $\xi$ with the help of an auxiliary random variable $\beta$ that takes values uniformly on $B$.  Consider the following transition kernel for moving in the augmented state space from $(\xi,\beta)$ to $(\xi',\beta')$:
\[
q(\xi',\beta'\cd\xi,\beta) \propto \pr(\xi'\cd\theta,A,B,H_p)I\left[\xi'\setminus \{\edge{\Pi_{A,\beta}(\xi')}{\beta}\}=\xi\setminus \{\edge{\Pi_{A,\beta}(\xi)}{\beta}\}\right],
\]
where $\pr(\xi'\cd\theta,A,B,H_p)$ is proportional to \cref{eq:HpDensityUnexpanded}.  This transition kernel allows transitions to any $\xi'\in\Xi(A,B)$ which differs from $\xi$ only in its match for $\beta$.  The states $\xi'$ which are accessible from the state $\xi$ are illustrated in \cref{fig:SwapsAllowed}.
We can move from any state $\xi$ to any other state $\xi'$ in at most $n_B$ steps, so the Markov chain with this transition kernel is irreducible.  Clearly it is also aperiodic and therefore ergodic.  Its stationary distribution is $ \pr(\xi\cd\theta,A,B,H_p)$ as desired. 

The densities of the allowed states have many terms in common.  By ignoring these common terms, we obtain from \cref{eq:fullconditionalofxib}
\[
q(\xi',\beta'\cd\xi,\beta)\propto \exp[w(a,\beta\cd\theta)- w\{a,\Pi_{B,a}(\xi)\cd\theta\}],
\]
where $a=\Pi_{A,\beta}(\xi')$ and $w$ is given by \cref{eq:xiproposal}.  Thus the proposal function can be computed very quickly, and it can be normalized over $\xi$ by simply summing over the permitted moves.  There are $n_A+1$ such moves, one for each possible value of $a\in A\cup\dummyelement$.  The full algorithm is described in \cref{alg:xiSampler}.

\begin{algorithm}[htbp]
\KwIn{Previous value $\xi$}
 $\beta\gets\sample{\mbox{Uniform over }B}$\;
 $\xi'\gets\xi\setminus\{\edge{\Pi_{A,\beta}(\xi)}{\beta}\}$ \# {remove the old match of $\beta$}\;
 $\alpha\gets\sample{\pr(a)\propto\exp[w(a,\beta\cd\theta)- w\{a, \Pi_{B,a}(\xi)\cd\theta\}]}$ for $a\in A\cup\dummyelement$\;
\If{$\alpha=\dummyelement$}{
\KwRet $\xi'$\;
}\Else{
\KwRet $\xi'\cup\{\edge{\alpha}{\beta}\}$\;
}
\caption{Sampler for $\xi$ using the auxiliary variable $\beta$.}
\label{alg:xiSampler}
\end{algorithm}

\end{document}